\RequirePackage{fix-cm}
\documentclass[smallextended]{svjour3}       
\smartqed  
\usepackage{amssymb}
\usepackage{amsmath}
\usepackage{graphicx}
\usepackage{color}
\journalname{Ricerche di Matematica. doi.org/ 10.1007/s11587-020-00512-w}
\begin{document}

\title{Rankine--Hugoniot conditions obtained by using the space--time Hamilton action}

\titlerunning{Hamilton's action  and shock waves}        
\author{Henri Gouin}

\institute{ \at
             Aix Marseille Univ, CNRS, IUSTI, UMR 7343, Marseille, France. \\
              \email{henri.gouin@univ-amu.fr; henri.gouin@yahoo.fr} \\ 
          ORCID iD: 0000-0003-4088-1386}
\date{Springer on line of the final version: 16 April 2020}

\maketitle
\noindent{\color{red}{Cite this article: Gouin, H. Rankine–Hugoniot conditions obtained by using the space–time Hamilton action. Ricerche mat (2020).\\ https://doi.org/10.1007/s11587-020-00512-w}}
\vskip 0.5cm

\begin{abstract}

In the \textit{quadri--dimensional   space--time}, the variation  of Hamilton's action  is a powerful tool to study the process equations for conservative  fluid media. In this framework, Hamilton's principle allows to  obtain   equation of motions,     equation of energy but also    Rankine-Hugoniot conditions.
The variational method may be a versatile key to obtain the shock-wave conditions for    complex media  when the equations of processes are not expressed by linear or quasi-linear differential equations.
\end{abstract}

\section{Introduction}

 In the literature, an ideal shock wave is mainly associated with hyperbolic systems of conservation laws  and represented by a moving surface dividing space in  two parts where a continuous solution exists, with a jump across this surface \cite{Dafermos}.
 The main problem of shocks   for hyperbolic systems was proposed by Riemann   but only in the case of one spatial dimension and  
shock solutions are weak solutions that must satisfy the conditions of Rankine--Hugoniot \cite{Krehl}. Although  shock waves are an important research topic for many years, the importance of the Riemann  problem  is not so clear in multidimensional situations.   In   case of complex medium, the Rankine--Hugoniot  conditions are often the subject of discussions 
\cite{Ruggeri}.
\\

The paper is destined to didactically present the  tool associated with variations of Hamilton's action in the physical quadri--dimensional   space--time (\textit{4-D space--time}) to  obtain the Rankine--Hugoniot conditions of shock waves for conservative (i.e non dissipative) media.\\ 
 The variation of  
Hamiton's action is related to the theory of distributions where a
decomposition theorem is associated with a    linear functional of \textit{virtual displacements} \cite{Serrin}. The virtual displacements, which are well-known in variational methods, are considered  as \textit{test
functions} whose supports are  compact manifolds \cite{Schwartz}.
The variation of Hamilton's action can be written in  a unique canonical zero order form   with respect both to the test functions and their  transverse (normal)
derivatives  to sub--manifolds  corresponding to  successive boundaries and edges \cite{Gouin1}.  \\
The equations of motion and  energy, and boundary conditions of   continuous media are deduced from Hamilton's principle : the motion is such that the action is zero for
	any virtual displacement    \cite{Casal}.  
When the Lagrangian  depends on
the strain tensor,  Hamilton's action   depends on the gradient of the virtual displacement  and leads  to existence of the Cauchy stress--tensor. When the Lagrangian  also depends on
over--strain tensor, then  the Hamilton  action   depends on  second gradient of the virtual displacement and we obtain     second--gradient media model   like for van der Waals fluids \cite{van der Waals,Germain}. 
\\
To understand the  proposed tool, we simply present the case of conservative  fluids (as  elastic media) and we prove that  Hamilton's principle  is   able to determine shock conditions  when  the variations of   Hamilton's action are associated with virtual displacements   in the  \textit{4-D} space--time. The method is able to determine  the Rankine--Hugoniot conditions as it was already performed for mixtures of fluids  \cite{Gouin}.\\ 
 Generally applied in relativity, the Hamilton method -- as commented in conclusion -- will be used    for  more  complex media,  when the equations of motions are not linear or quasi-linear,  in a forthcoming article \cite{Gavrilyuk2} ; they are cases when the Rankine--Hugoniot conditions cannot be  obtained by classical methods associated with problems of hyperbolic equations of motions as in \cite{Lax}.
 \\
 
 The paper is presented as follows :\\ 
In Section 2,  Hamilton's action and Hamilton's principle are developed in the   \textit{4-D}   space--time.\\  In section 3,  fluids (and more generally elastic media) are considered, and  two forms of virtual displacements  associated with variations of Hamilton's action in  \textit{4-D}   space--time and \textit{4-D} reference--space are analyzed and compared.  The form of virtual displacements in    space--time  allows to obtain  the equations of motion  and  energy; the form in     reference--space allows to obtain   the   equation  of motion  in   thermodynamic form and the specific entropy conservation along trajectories.\\ Section 4 introduces the moving surfaces  of discontinuity. \\
 In Section 5, the   virtual displacements in the   $\textit{4-D}$   space--time yield the complete set of Rankine--Hugoniot conditions; it is not the same for  virtual displacements in the   $\textit{4-D}$  reference--space.\\    In  
 conclusion, the  virtual displacements in the   $\textit{4-D}$  space--time are highlighted.

\section{The Hamilton principle}
\subsection{The Hamilton action}
\quad\,\ In the  \textit{4-D}  space--time $\mathcal W$, we consider a continuous medium  
of position variables $\boldsymbol{z}=\left(
\begin{array}{c}
t \\
\boldsymbol{x}
\end{array}
\right) \equiv \left\{ z^{i}\right\},\,$ $(i=0,1,2,3)$, where $t$ is the time  and $%
\boldsymbol{x}\equiv \left\{ x^{i}\right\},\, (i=1,2,3)$ denote the Euler  variables; we also write :
\begin{equation*}
\boldsymbol{z}^\star =\left(
z^0,
z^1,  
z^2,  
z^3  
\right)  \quad \text{where}\quad z^0=t \, \ \text{and}\,\ z^1=x^1,\, z^2 = x^2,\, z^3 = x^3
\end{equation*}

We   also consider  the most general $\textit{4-D}$ reference--space $\mathcal W_0$ of position variables $\boldsymbol{Z}=\left(
\begin{array}{c}
\lambda \\
\boldsymbol{X}
\end{array}
\right) \equiv \left\{ Z^{i}\right\}, \,$ $(i=0,1,2,3)$,  where $\lambda$ is a real scalar parameter  and $%
\boldsymbol{X}\equiv \left\{ X^{i}\right\}, \,(i=1,2,3)$ denote  the Lagrange variables;  
we also write :
\begin{equation*}
\boldsymbol{Z}^\star =\left(
Z^1,
Z^2,  
Z^3,  
Z^4  
\right)  \quad \text{where}\quad Z^0=\lambda \, \ \text{and}\,\ Z^1=X^1,\, Z^2 = X^2,\, Z^3 = X^3
\end{equation*}
A   continous--medium motion is represented by the mapping :
\begin{equation}
\boldsymbol{z} =  \mathbf\Phi (\boldsymbol{Z})\label{motion}
\end{equation}
{The Hamilton  action $a$ of an elastic medium} is expressed as \cite{Casal,Seliger-Witham} :
\begin{equation*}
a = \int_{\mathcal W} L\, dw \quad \text{with}\quad L =\mathcal{L}\left(\boldsymbol z, \boldsymbol Z, \frac{\partial \boldsymbol z}{\partial \boldsymbol Z}\right)
\end{equation*}
where $dw= dv\times dt$ is the volume-time measure in the  \textit{4-D}  space--time $\mathcal W$, $L$ is the Lagrangian,   and  
\begin{equation}
\mathcal B = \frac{\partial \boldsymbol z}{\partial \boldsymbol Z}\equiv  \frac{\partial\mathbf\Phi\left(\boldsymbol Z\right)}{\partial \boldsymbol Z}\quad\text{where}\quad \frac{\partial \boldsymbol z}{\partial \boldsymbol Z}\equiv\left\{\frac{\partial   z^i}{\partial  Z^j}\right\}\quad\text{with}\quad i,  j\in\left\{0,1,2,3\right\}\label{B1}
\end{equation}
denotes the  tangent linear application  of  motion $\mathbf\Phi$ in the  \textit{4-D}  space--time (\footnote{We notice that $L$ has the dimension of an energy per unit volume and consequently, $a=\int_{\mathcal W} L\, dw$ has the dimension of an action.}).
\subsection{Variation of the Hamilton  action}

To vary a motion of the medium, we consider a family $\mathbf\Psi$ associated with a real parameter $\varepsilon$ belonging to the vicinity of $0$ \cite{Serrin,Casal,Gavrilyuk} :
\begin{equation} 
\boldsymbol{z} =  \mathbf\Psi (\boldsymbol{Z},\varepsilon)  \quad\text{such that}\quad  \mathbf\Psi (\boldsymbol{Z},0) =   \mathbf\Phi (\boldsymbol{Z})  \label{var}      
 \end{equation}
 Then, 
\begin{equation*}
a =f(\varepsilon) \  \text{and the variation}\,\ \delta a\,\  \text{is defined by}\  \delta a =f^\prime(0),
 \end{equation*}
 where  for a variation, the differential  is denoted $\delta$ in place of $d$. 
 Two possibilities can be considered to obtain the variation of action $a$.
 From 
 \begin{equation}
 \delta \boldsymbol z =  \frac{\partial\mathbf\Psi}{\partial \boldsymbol Z}\,  \delta \boldsymbol Z+ \frac{\partial\mathbf\Psi}{\partial \varepsilon} \, \delta\varepsilon\quad \text{with}\quad \delta\varepsilon=1,\quad  \text{we deduce :} \label{variationdiff}
 \end{equation}

  $\bullet$\quad  A first variation :    \begin{equation*}\delta \boldsymbol z=   \boldsymbol{\tilde\zeta} \,\quad {\rm when}\,\ \delta \boldsymbol Z =\boldsymbol 0
 \end{equation*}                     
             
  $\bullet$\quad  A second variation :    \begin{equation*}\delta \boldsymbol Z =\boldsymbol{\hat\zeta}  \,\quad {\rm when}\,\ \delta \boldsymbol z =\boldsymbol 0
  \end{equation*}
   where  symbols \textit{tilde} and \textit{hat} respectively  denote the first and second variations associated with \eqref{var}.
   \\
    \begin{remark}:
   	The two variations respectively denoted by $\tilde \delta$ and $\hat\delta$ are not independent.
   	In fact   \eqref{var} and \eqref{variationdiff}  imply :
   	$$\displaystyle
   	\frac{\partial  \mathbf\Psi(\boldsymbol Z,0)}{\partial \varepsilon}+\frac{\partial  \mathbf\Psi(\boldsymbol Z,0)}{\partial \boldsymbol Z}\, \boldsymbol{\hat\zeta} =\boldsymbol 0
   	$$
   	which can be written :
   	\begin{equation}
   	\boldsymbol{\tilde\zeta} +\mathcal B\, \boldsymbol{\hat\zeta}=\boldsymbol 0\label{relvirtualdsipl}
   	\end{equation}          	
   \end{remark}
We use the  notations (\footnote{For  vectors $\boldsymbol{a}$ and $\boldsymbol{b}$,   $\boldsymbol{a}%
	^{\star }\boldsymbol{b}$, where superscript $^\star$ denotes the transposition, is the scalar product (line vector
	$\boldsymbol{a}^{\star}$
	is multiplied by column vector $\boldsymbol{b}$); for the sake of simplicity, we also denote $\boldsymbol{a}%
	^{\star }\boldsymbol{a}= \boldsymbol{a}^2$. Tensor $\boldsymbol{a} {%
		\ }\boldsymbol{b}^{\star}$ (or $\boldsymbol{a}\otimes \boldsymbol{b}$) is
	the product of column vector $\boldsymbol{a}$ by line vector $\boldsymbol{b}%
	^{\star}$. Tensor $\boldsymbol{%
		1}$ is the identity.  In the physical \textit{3-D} space,  we denote  the zero matrix   by  $\boldsymbol{O}$  and  the zero vector  
	by $\boldsymbol{0}$, respectively.}) : \\ Operators $\text{div}$ and $\text{Div}$,
 $\nabla= {\rm grad}=\left(\dfrac{\partial}{\partial\boldsymbol{x}}\right)^\star$ and  $\text{Grad}=\left(\dfrac{\partial}{\partial\boldsymbol{z}}\right)^\star$, are  the divergence  and gradient
   in the \textit{3-D} and \textit{4-D} physical spaces, respectively. \\
   The divergence of
   a second order tensor $\mathcal A$ is a covector (i.e. a form) defined as :
   \begin{equation*}
   \text{Div}\left( \mathcal A\,\boldsymbol{h_0}\right) =\text{Div}\left(\mathcal A\right) \boldsymbol{h_0}
   \end{equation*}
   where $\boldsymbol{h_0}$ is a constant vector field  in the    \textit{4-D} space-time. In particular,
   for any linear transformation $\mathcal A$ and any vector
   field $\boldsymbol{h}$, we get : 
   \begin{equation*}
   \text{Div}(\mathcal A\,\boldsymbol{h})=
   (\text{Div }\mathcal A)\,\boldsymbol{h}+ \text{Tr}\left(\mathcal A\,\dfrac{\partial
   	\boldsymbol{h}}{\partial \boldsymbol{z}}\right)
   \end{equation*}
   where Tr denotes the trace operator.  
   If $f(\mathcal A)$ is any scalar function  of $\mathcal A$, we denote :
   \begin{equation*}
   \nabla_{\mathcal A} f = \left(\frac{\partial f}{\partial \mathcal A}\right)^\star  \quad\text{with}\quad\left(\frac{\partial f}{\partial \mathcal A}\right)_j^i=\left(\frac{\partial f}{\partial \mathcal A_i^j}\right)
   \end{equation*}
   where $i$ is the line index and $j$ the column index of $\mathcal A$ and we denote :
   \begin{equation*}
   df(\mathcal A) = \nabla_{\mathcal A} f: d\mathcal A \equiv \left(\frac{\partial f}{\partial \mathcal A}\right)_j^i\, d\mathcal A_i^j
   \end{equation*}  
   where repeated indices correspond to the summation.\\
     The notations  are similar for \textit{4-D} reference space $\mathcal W_0$; in this case, we simply add the   subscript ${0}$. \\
    
  Equation \eqref{motion} allows  to write the variations of Hamilton's action $a$ as well in variable     $\boldsymbol Z$ than in variable $\boldsymbol z$:  
 \begin{center}
  	\begin{tabular}{|c|c|c}
  		\hline
  		&
  		\\
  		\multicolumn{1}{|c|}{$ \delta \boldsymbol z=   \boldsymbol {\tilde\zeta},\,\ \delta \boldsymbol Z = \boldsymbol 0 \quad \Longrightarrow \quad \displaystyle \tilde\delta\mathcal B = \frac{\partial \boldsymbol {\tilde\zeta}}{\partial \boldsymbol z}\, \mathcal B$ } & $ \delta \boldsymbol Z =\boldsymbol{\hat\zeta}, \,\ \delta \boldsymbol z= \boldsymbol  0\quad       \Longrightarrow \quad \hat\delta\mathcal B = -\displaystyle\mathcal B\,\frac{\partial \boldsymbol {\hat\zeta}}{\partial \boldsymbol Z}$ 
  		 \\
  		&
  		 \\
  		\hline 
  		&
  	      \\
  		\multicolumn{1}{|c|}{$\displaystyle a= \int_{\mathcal W_0}  L\, {\rm det}\,\mathcal B\,dw_0$} &  {$\displaystyle a= \int_{\mathcal W}  L\,  \,dw$}  
  		\\
  		&
  		\\
  		$\displaystyle\delta a =\int_{\mathcal W_0}\left\{\frac{\partial L}{\partial\boldsymbol z}\,{\rm det}\, \mathcal B\,\boldsymbol {\tilde\zeta}+{\rm Tr}\left(\frac{\partial (L\,{\rm det}\,\mathcal B) }{\partial\mathcal B }\,\tilde\delta \mathcal B\right)\right\} dw_0$ & 	$\displaystyle\delta a =\int_{\mathcal W}\left\{\frac{\partial L}{\partial\boldsymbol Z} \, \hat\zeta+{\rm Tr}\left(\frac{\partial L}{\partial\mathcal B }\,\hat\delta \mathcal B\right)\right\} dw$  \\ 
&  
	\\ 
  		 	\multicolumn{1}{|c|} {$\Longrightarrow \quad  \displaystyle \delta a= \int_{\mathcal W} \{ \mathcal F^\star\, \boldsymbol{\tilde\zeta}+ {\rm Tr}\,\left(T\,\frac{\partial\boldsymbol{\tilde\zeta}}{\partial\boldsymbol z}\right)dw = $}  &  {$\Longrightarrow \quad  \displaystyle \delta a= \int_{\mathcal W_0} \{ \mathcal F_0^\star\, \boldsymbol{\hat\zeta}+ {\rm Tr}\,\left(T_0\,\frac{\partial\boldsymbol{\hat\zeta}}{\partial\boldsymbol Z}\right)dw_0 = $}  
  		\\ 
  		&
  		\\
  		$\displaystyle\int_{\mathcal W} \left\{ \mathcal F^\star- {\rm Div}\,
  		T\right\}\boldsymbol{\tilde\zeta}\,dw+\int_{\partial \mathcal W}\boldsymbol N^\star T\,\boldsymbol{\tilde\zeta}\, d\sigma$&$\displaystyle\int_{\mathcal W_0} \left\{ \mathcal F_0^\star- {\rm Div_0}\,
  		T_0\right\}\boldsymbol{\hat\zeta}\,dw_0+\int_{\partial \mathcal W_0}\boldsymbol N_0^\star T_0\,\boldsymbol{\hat\zeta}\, d\sigma_0$
  		\\ 
  		&
  		\\
  		$\displaystyle  {\rm with}\quad \mathcal F^\star =\frac{\partial L}{\partial\boldsymbol z}  \quad{\rm and}\quad T = L\, \boldsymbol 1+ \mathcal B\,\frac{\partial L}{\partial\mathcal B}$
  		&  	
  		$\displaystyle  {\rm with}\ \ \mathcal F_0^\star ={\rm det}\,\mathcal B\,\frac{\partial L}{\partial\boldsymbol Z}  \ \ {\rm and}\ \ T_0 = -{\rm det}\,\mathcal B\, \frac{\partial L}{\partial\mathcal B}\,\mathcal B$\\
  		&
  		\\
  		\hline     
  		
  	\end{tabular}\vskip0.5cm
  
  \end{center}

\noindent In relativity,  $T$ is called    energy-impulsion tensor (see \cite{Souriau}) and $\mathcal F$ is the extended force;  Div $T$ is a form of $\mathcal W$;\,\ det\,\ denotes the determinant.
The boundary of $\mathcal W$  is denoted $\partial\mathcal W$ with measure $d\sigma$; $\boldsymbol{N}^\star$ denotes the linear form such that $\boldsymbol{N}^\star\boldsymbol{\tilde\zeta}\,d\sigma= \rm{det}\left(\boldsymbol{\tilde\zeta},d_1\boldsymbol{z},d_2\boldsymbol{z},d_3\boldsymbol{z}\right)$, where $d_1\boldsymbol{z},d_2\boldsymbol{z},d_3\boldsymbol{z}$ are the differentials associated with  coordinate lines of $\partial\mathcal W$ in the  \textit{4-D} space--time. Similar notations are  used for $\mathcal W_0$ and its boundary  $\partial\mathcal W_0$ with the additional subscript ${0}$, but $T_0$ and $\mathcal F_0$ don't have  any names.\\ Thanks to the Stokes formula,
integrals $\displaystyle\int_{\partial \mathcal W}\boldsymbol N^\star T\,\boldsymbol{\tilde\zeta}\, d\sigma$ and $\displaystyle\int_{\partial \mathcal W_0}\boldsymbol N_0^\star T_0\,\boldsymbol{\hat\zeta}\, d\sigma_0$ correspond to the integration of ${\rm Div}\, (T\,\boldsymbol{ \tilde\zeta}) $ and  ${\rm Div}_0\, (T_0\,\boldsymbol{ \hat\zeta}) $ on boundaries $\partial \mathcal W$ and $\partial \mathcal W_0$, respectively.
            
          \section{Motions of fluid media}
            From  \eqref{B1}, we note  that $\mathcal B$ can be written :
          \begin{equation}
          \mathcal B=\left(
          \begin{array}{cc}
          \mu  &\,\ \boldsymbol w^\star  \\
          \boldsymbol r  &\  \boldsymbol  B     
          \end{array}%
          \right)\,\ \rm {corresponding\ to\ the\   relations }\quad\left\{
          \begin{array}{l}
          \displaystyle
          dt=\mu \,d\lambda+ \boldsymbol{w}^\star d\boldsymbol{X} \\
          \\
          \displaystyle d\boldsymbol{x}= \boldsymbol{r}\,d\lambda +\boldsymbol{B}\, d\boldsymbol{X}
          \end{array}%
          \right.\label{B2}
          \end{equation}
          where $\mu$ is a scalar, $\boldsymbol{w}^\star$ a form, $\boldsymbol{r}$ a vector and $\boldsymbol{B}$ is a \textit{3-D} linear application.
          Eliminating $d\lambda$ between the two  relations of system \eqref{B2}, we obtain :
          \begin{equation}
          \boldsymbol{v}\equiv \frac{\partial \boldsymbol{x}(\boldsymbol{X},t)}{\partial t}\equiv\frac{\boldsymbol{r}}{\mu} \quad\text{and}\quad \boldsymbol{F}\equiv \frac{\partial \boldsymbol{x}(\boldsymbol{X},t)}{\partial \boldsymbol{X}}= \boldsymbol{B}- \frac{\boldsymbol{r}}{\mu} \, \boldsymbol{w}^\star= \boldsymbol{B}-  \boldsymbol{v}   \, \boldsymbol{w}^\star \label{keyF}
          \end{equation}
          where $\boldsymbol{v}$ is the fluid velocity and $\boldsymbol{F}$ denotes the  tangent linear deformation of the medium. \\ In the particular case when we choose $\lambda =t$, we obtain :
          \begin{equation*}
          \mathcal A\equiv \frac{\partial\boldsymbol z}{\partial\boldsymbol Z_0}=    \left(
          \begin{array}{cc}
          1  &\ \boldsymbol 0^\star  \\
          \boldsymbol v  &\  \boldsymbol  F     
          \end{array}%
          \right)\quad \rm{where}\quad\boldsymbol Z_0=
          \left(
          \begin{array}{c}
          t \\
          \boldsymbol{X}
          \end{array}
          \right)
          \end{equation*}
          and we denote  $\mathcal W^\prime_0$ the associated \textit{4-D} reference space of coordinates $(t,\boldsymbol{X})$. Second order tensors   $\mathcal A$  and $\mathcal B $ are connected by the relation:
          \begin{equation*}
          \mathcal B = \mathcal A\, \boldsymbol\Lambda\quad{\rm where}\ \quad \boldsymbol \Lambda \equiv \frac{\partial\boldsymbol Z_0}{\partial\boldsymbol Z}=    \left(
          \begin{array}{cc}
          \mu &\ \boldsymbol w^\star  \\
          \boldsymbol 0 &\  \boldsymbol  1     
          \end{array}%
          \right)
          \end{equation*}  
          with $ dt(\lambda,\boldsymbol{X})=\mu \,d\lambda+ \boldsymbol{w}^\star d\boldsymbol{X}$\ and\ $d\boldsymbol{X}(\lambda,\boldsymbol{X})=\boldsymbol 0\, d\lambda+\boldsymbol{1}\, d\boldsymbol{X}$, which implies:
          \begin{equation*}
          \mu = \frac{\partial t(\lambda,\boldsymbol X)}{\partial  \lambda}\quad\text{and}\quad\boldsymbol w^\star =\frac{\partial t(\lambda,\boldsymbol X)}{\partial  \boldsymbol X}
          \end{equation*}
          \subsection{The Lagrangian of a fluid and its consequences}
          
         If we denote $\displaystyle  \mathcal V =\left(
          \begin{array}{c}
          1\\
          \boldsymbol{v}
          \end{array}
          \right)$ the \textit{4-D} velocity field in space--time associated with the material derivative of   $\displaystyle  \boldsymbol z =\left(
          \begin{array}{c}
          t \\
          \boldsymbol{x}
          \end{array}
          \right)$,
        conservative fluids are elastic media with Lagrangian. The Lagrangian is a function of $\boldsymbol z$,  $\boldsymbol Z$ and $\mathcal B$ such that \cite{Lamb,Truesdel1} (\footnote {Let us note that $\displaystyle \frac{1}{2} \rho\, \mathcal V^2 =\frac{1}{2}\,\rho\,\left(1+\boldsymbol v^2\right)$. Because potential energy $\Omega$ is only defined to within an arbitrary additive constant,  the term $\displaystyle\frac{1}{2}\,\rho\times 1$ where  $1$ has the physical dimension of a velocity square  can be added to  $\rho\,\Omega(\boldsymbol z)$, without changing the Hamilton action.   Consequently, the Lagrangian can also be written
            $\displaystyle L = \frac{1}{2}\, \rho\,\boldsymbol v^2 
          -\rho \,\alpha(\rho,s) -\rho \,\Omega(\boldsymbol z)	$.}):  
           \begin{equation*}
          L = \frac{1}{2}\, \rho\,\mathcal V^2 
          -\rho \,\alpha(\rho,s) -\rho \,\Omega(\boldsymbol z) \end{equation*} 
 where : \\
          
          $\bullet$ The volume kinetic energy is\,\ $\displaystyle \frac{1}{2}\,\rho\,\boldsymbol v^2$,   $\rho$ being the mass density. 
        \begin{equation}
       \rm {In\ fact},\quad \frac{1}{2}\,  \mathcal V^2 \equiv \frac{1}{2 }\,  \frac{\boldsymbol I^\star \mathcal B^\star\mathcal B\,\boldsymbol I}{\mu^2} \quad \rm{where}\ \,\boldsymbol I = \left(
       \begin{array}{c}
       {1} \\
       \boldsymbol{0}
       \end{array}
       \right) \ {\rm and}\ \mu = \boldsymbol I^\star \mathcal  B\,\boldsymbol I \label{key}
          \end{equation} 
          
           $\bullet$ The specific internal energy $\alpha(\rho,s)$ is function of the mass  density $\rho$ and the specific entropy $s$.  Due to the mass conservation, the image of the density in the reference space $\mathcal W_0$ is conserved. For conservative motions, the entropy $s$ is attached to the reference space. Then :
           \begin{equation*}
            \frac{ \rho\ \rm {det} \,\mathcal B}{\mu}  =  f(\boldsymbol X)\qquad\text{and}\qquad s= g(\boldsymbol Z)
           \end{equation*}
         where $f$ and $g$ are two real scalar functions defined in Lagrange variables. \\
       		 
             $\bullet$ The specific  potential  energy $\Omega$ due to  body forces is defined in $\mathcal W$ : \begin{equation*}
             \Omega =\Omega(\boldsymbol z)
           \end{equation*} 
           We obtain :
              \begin{equation*}
              \frac{\partial L}{\partial \rho}= \frac{1}{2}\, \mathcal {V}^2 -h-\Omega\equiv m,\quad \frac{\partial L}{\partial s}=-\rho\,\theta\quad\rm{and} \quad\frac{\partial L}{\partial \left(\displaystyle\frac{1}{2}\,\mathcal {V}^2\right)}=\rho
              \end{equation*}
              where $\displaystyle h=\alpha+\rho\,\frac{\partial \alpha(\rho,s)}{\partial \rho}$ is the specific enthalpy, $\theta$ the Kelvin temperature and   $\displaystyle
              p=\rho^2\frac{\partial \alpha}{\partial \rho}$ the pressure  of the fluid.\\ Let us calculate $\displaystyle \frac{\partial L}{\partial \mathcal B }$\ or firstly\ $\displaystyle \frac{\partial \rho}{\partial \mathcal B }$\ and\ $\displaystyle\,\frac{\partial \left(\displaystyle \frac{1}{2}\, \mathcal {V}^2\right)}{\partial\mathcal B}$.\\
              
               \noindent From
              $d(\rm{det}\,\mathcal B) =\rm{det}\,\mathcal B\  \text{Tr}\left(\mathcal B^{-1}d\mathcal B\right)$,\  $d\mu = \boldsymbol I^\star d\mathcal \mathcal B\,\boldsymbol I =\rm{Tr}\left(\boldsymbol I\,\boldsymbol I^\star d\mathcal B\right)$ \ and  \eqref{key},  we obtain the three relations :
               \begin{equation*}
            \frac{\partial \rho}{\partial \mathcal B }=f(\boldsymbol{X})\,\mu\,\frac{\partial}{\partial\mathcal B}\left(\frac{1}{\rm{det}\,\mathcal B}\right)+ f(\boldsymbol{X})\,\frac{1}{\rm{det}\,\mathcal B}\, \frac{\partial\mu}{\partial\mathcal B}=
            -\frac{f(\boldsymbol{X})\,\mu}{\rm{det}\,\mathcal B}\, \mathcal B^{-1}+f(\boldsymbol{X})\frac{\boldsymbol I\,\boldsymbol I^\star}{\rm{det}\,\mathcal B} 
             \end{equation*}
             \begin{equation*}
             {\rm{det}\,\mathcal B}\,\frac{\partial \rho}{\partial\mathcal B } \, { \mathcal B }=-f(\boldsymbol{X})\,\mu\, \left(
             \begin{array}{cc}
             1 &\ \boldsymbol 0^\star  \\
             \boldsymbol 0 &\  \boldsymbol  1     
             \end{array}%
             \right)+ f(\boldsymbol{X})\, \left(
             \begin{array}{cc}
             \mu &\ \boldsymbol w^\star  \\
             \boldsymbol 0 &\  \boldsymbol  O    
             \end{array}%
             \right)=f(\boldsymbol{X})\, \left(
             \begin{array}{cc}
             0 &\ \boldsymbol w^\star  \\
             \boldsymbol 0 &\,-\mu  \boldsymbol  1     
             \end{array}%
             \right) \label{keydet}
             \end{equation*}
           \begin{equation*}
           \mathcal B\,\frac{\partial \rho}{\partial \mathcal B } =-\rho\,\left(\boldsymbol{1}-  \,\mathcal{V}\boldsymbol{I}^\star\right),\quad   \frac{\partial \left(\displaystyle\frac{1}{2}\, \mathcal {V}^2\right)}{\partial\mathcal B}= \frac{\boldsymbol I\,\boldsymbol I^\star}{\mu}\left(\frac{\mathcal B^\star}{\mu}- \frac{1}{2}\, \mathcal {V}^2\,\boldsymbol 1\right)
             \end{equation*}
              Moreover, from \eqref{keyF}, we  obtain the three relations :
             \begin{equation*}
        \boldsymbol I\,\boldsymbol I^\star=
       \left(
       \begin{array}{cc}
       1 &\ \boldsymbol 0^\star  \\
       \boldsymbol 0 &\  \boldsymbol  O     
       \end{array}%
       \right)  \Longrightarrow\quad {\rm{det}\ \mathcal B}\,\frac{\partial \left(\displaystyle\frac{1}{2}\, \mathcal {V}^2\right)}{\partial\mathcal B}\, \mathcal B  =  \frac{{\rm{det}\,\mathcal B}}{\mu}\, \left(
          \begin{array}{cc}
          0 &\ \boldsymbol v^\star \boldsymbol  F     
         \\
          \boldsymbol 0 &\   \boldsymbol  O     
          \end{array}%
          \right)\label{keyB}
             \end{equation*}
              \begin{equation*}
   \mathcal B \, \,\frac{\partial \left(\displaystyle\frac{1}{2}\, \mathcal {V}^2\right)}{\partial\mathcal B}= \left(
              \begin{array}{cc}
              -\boldsymbol v^2&\ \boldsymbol v^\star      
              \\
              -(\boldsymbol v^2)\,\boldsymbol v\ &\   \boldsymbol v\boldsymbol v ^\star
              \end{array}%
              \right)= \mathcal V\mathcal V^\star \left(\boldsymbol 1- \mathcal V\boldsymbol I^\star\right)
            \end{equation*}
             \begin{equation*}
             \frac{\partial L}{\partial\boldsymbol Z}=\left(\frac{1}{2}\, \mathcal {V}^2-\alpha -\rho\,\frac{\partial\alpha}{\partial\rho}- {\Omega}\right)\,\frac{\partial\rho}{\partial\boldsymbol Z}-\rho\,\frac{\partial\alpha}{\partial s}\, \frac{\partial s}{\partial\boldsymbol Z}\ 
             \end{equation*}
             Additively, we obtain :
             \begin{equation*}
             \frac{\partial L}{\partial\mathcal B} =\rho\,\frac{\partial\left(\displaystyle\frac{1}{2}\, \mathcal {V}^2\right)}{\partial \mathcal B}+  \left(\displaystyle
             \frac{1}{2}\, \mathcal {V}^2-\alpha-\rho\, \frac{\partial\alpha}{\partial\rho}-\Omega\right) \,\frac{\partial\rho}{\partial \mathcal B}\quad\Longrightarrow
             \end{equation*}
             \begin{equation*}
             T_0=-\,\text{det}\,\mathcal B\, \left(\rho\,\frac{\partial\left(\displaystyle\frac{1}{2}\, \mathcal {V}^2\right)}{\partial \mathcal B}+ m\,\frac{\partial\rho}{\partial \mathcal B}\right)\, \mathcal B 
             \end{equation*} 
             \subsection{ The Hamilton  principle}
            {\bf  Principle} \cite{Casal,Gavrilyuk}:  \textit{For  all virtual displacements which are  null on the  boundary   $\partial\mathcal W$,  (respectively on the boundary $\partial\mathcal W_0$), the variations of   Hamilton's action are zero.}\\

          \noindent  From Hamilton's principle and  calculations in Section 3.1, we obtain :
             \begin{center}
             	\begin{tabular}{|c|c|c}
             		\hline
             		&
             		\\
             		\multicolumn{1}{|c|}{$\displaystyle \mathcal F^\star =-\rho\,\frac{\partial \Omega}{\partial\boldsymbol z},  \ T =  \left(
             			\begin{array}{cc}
             			-e &\rho\, \boldsymbol v^\star  \\
             			-\left(e+p\right)  \boldsymbol v &\ \, \rho\,\boldsymbol v \boldsymbol v^\star+p\boldsymbol 1
             			\end{array}%
             			\right)$} &  {$\displaystyle \mathcal F_0^\star =-\mu\,f(\boldsymbol X) \, \theta\, \frac{\partial s}{\partial\boldsymbol Z} +\mu\,m \left(0, \frac{\partial f}{\partial X}\right) $}
             		\\&\\
             	 $e=\displaystyle\rho\left(\frac{1}{2}\,\boldsymbol v^2+ \alpha+\Omega\right),\,\  p =\displaystyle \rho^2\,\frac{\partial \alpha(\rho,s)}{\partial\rho}$ 
             		&
             		{$T_0 = f(\boldsymbol X)\, \left(
             		\begin{array}{cc}
             			0\ \ &- \boldsymbol v^\star\boldsymbol F - m\,\boldsymbol w^\star  \\
             			  \boldsymbol 0\ \ &\  m\,\mu\,1
             		\end{array}%
             		\right)$}
             		\\
             		&
             		\\
             		\hline 
             		&
             		\\
             		\multicolumn{1}{|c|}{ We obtain:
             		$\mathcal F^\star -{\text{Div}}\ T=\boldsymbol{ 0}^\star$\quad$\Longleftrightarrow$}
             	 & {We obtain: $  \mathcal F_0^\star -{\text{Div}_0}\ T_0=\boldsymbol{ 0}^\star$}\quad$\Longleftrightarrow$
             			\\  &
             			\\
             		\multicolumn{1}{|c|} {
             			 $ \displaystyle  \frac{\partial e}{\partial t}+{\rm{div}}((e+p)\boldsymbol v)-\rho\,\frac{\partial \Omega}{\partial t}=0$,} &  { $\displaystyle \mu\,f(\boldsymbol{X})\,\theta \, \frac{\partial s}{\partial\lambda}=0,   $}  
             		\\ 
             		&
             		\\
             		 $  \displaystyle\frac{\partial\rho\boldsymbol v^\star}{\partial t}+{\rm{div}}\left( \rho\,\boldsymbol v \boldsymbol v^\star+p\boldsymbol 1\right)+\rho\,\frac{\partial\Omega}{\partial\boldsymbol x}=\boldsymbol 0^\star$ 
             		&  	
             	 $ \displaystyle\frac{\partial }{\partial\lambda}\left(\boldsymbol{v}^\star \boldsymbol{F}+m\,\boldsymbol{w}^\star\right)=\mu\,\theta\, \frac{\partial s}{\partial\boldsymbol X}+ \frac{\partial(\mu m) }{\partial\boldsymbol X}$\\
             		&
             		\\
             		\hline   
             	\end{tabular}
             \end{center}
         \vskip 0.5cm
 In the first column, $e$ is the total volume energy of the fluid ; we firstly obtain the classical \textit{equation of energy} and secondly  the \textit{equation of motions}.\\
         In the second column, although it is not necessary, we can choose the parameter $\lambda =t$ and we obtain $\mu =1$,  $\boldsymbol w= \boldsymbol 0$. Consequently  in Lagrange variables, we get : 
         \begin{equation}
         \frac{\partial s(t,\boldsymbol X)}{\partial t} =0\quad \Longleftrightarrow\quad\dot s =0\qquad \text{and}\qquad\frac{\partial \boldsymbol v^\star\boldsymbol{F} }{\partial t}=\theta\,\frac{\partial s}{\partial \boldsymbol X}+\frac{\partial m}{\partial \boldsymbol X}\label{refeq}       
         \end{equation}
         The first equation \eqref{refeq}$^1$  is the conservation of the specific entropy   along   fluid trajectories.
       Due to 
       \begin{equation*}
       \frac{\partial \boldsymbol v^\star\boldsymbol{F} }{\partial t} = \boldsymbol a^\star\boldsymbol{F}+  \frac{1}{2}\frac{\partial \boldsymbol v^2 }{\partial  \boldsymbol x}\, \boldsymbol{F}
       \end{equation*}
         where $\boldsymbol a$ is the acceleration vector,  the second equation \eqref{refeq}$^2$ writes in variables $(t, \boldsymbol X)$ : 
       $$\left(\displaystyle  \boldsymbol a^\star+\frac{\partial \left(h+\Omega\right)}
         {\partial x}-\theta\,\frac{\partial s}{\partial \boldsymbol x}\right)\,\boldsymbol{F}=\boldsymbol{0}^\star$$
       and finally :
             \begin{equation*}
             \boldsymbol a +{\rm grad}{\, \left(h+\Omega\right)}
        -\theta\,{\rm grad}{\,  s}=\boldsymbol{0}
         \end{equation*}
         This   equation is a thermodynamic form of the equation of motion \cite{Serrin}.
         \section{Moving surface}
         \subsection{Generality \cite{Hadamard}}
        In   \eqref{motion}, one can choose  the representation $t= \ell  (\lambda,\boldsymbol{X})$, and consequently its derivatives 
       \begin{equation*}
       \mu =   \frac{\partial \ell  (\lambda,\boldsymbol{X})}{\partial \lambda}\quad\text{and}\quad  \boldsymbol w^\star = \frac{\partial \ell  (\lambda,\boldsymbol{X})}{\partial \boldsymbol X}\end{equation*}  such that the image  by  application $\mathbf\Phi^{-1}$ of a moving surface $\Sigma$  (which is a \textit{3-D} manifold of \textit{4-D} space--time $\mathcal W$) is the hyperplan $\Sigma_0$ of equation $\lambda =0$.  The surface $\Sigma_0$   can be considered as  a  \textit{3-D} material--manifold of $\mathcal W_0^\prime$  with equation $t=\ell (0,\boldsymbol X)$.  
       Along $\Sigma_0$ we obtain :
       $$
        dt = \boldsymbol w^\star d\boldsymbol X  
        $$
        and we can write :
       \begin{equation}
       \boldsymbol N_0^\star\, d\boldsymbol Z_0=0\quad\text{with}\quad \boldsymbol N_0^\star=\pi  \left(-1, \boldsymbol w^\star\right) \label{normal0}
       \end{equation}
       where $\pi$ is a coefficient of proportionality. Along $\Sigma$,   we also have:
    \begin{equation}
        \boldsymbol N^\star d\boldsymbol z=0\quad \text{with}\quad \boldsymbol{ N}^\star= \left(-\it{D_n}, \boldsymbol n^\star\right)\label{normal}
     \end{equation}
       where $\boldsymbol n$ is the unit normal vector to $S_t$ section of $\Sigma$ at time $t$ and $\it{D_n}$ is the velocity of $S_t$. From \eqref{normal}, we deduce $  \boldsymbol N^\star \mathcal A\, d\boldsymbol Z_0=0$. Then, from \eqref{normal0} 
       \begin{equation}
      \boldsymbol N^\star \mathcal A= \kappa\,\boldsymbol N_0^\star \quad\text{with}\quad \boldsymbol N^\star\mathcal A  =\left(u, \boldsymbol n^\star\boldsymbol F\right)\label{keyN}
       \end{equation}
       where $u=\boldsymbol n^\star \boldsymbol v-\it{D_n}$ is the medium velocity with respect to the moving surface  $\Sigma$ and $\kappa$ is a Lagrange multiplier. 
       By similarity with $  \boldsymbol N$, we can write :  \begin{equation*}
       \boldsymbol N_0^\star=   \left(-{\it D_{n_0}}, \boldsymbol n_0^\star\right) \quad \text{and consequently}\quad \boldsymbol w^\star =\frac{\boldsymbol n_{0}^\star}{{\it D_{n_0}}} 
        \end{equation*}
     where ${\it D_{n_0}}$ is the velocity of $S_0$ image of $S_t$ in the  \textit{3-D} reference--space of variables $\boldsymbol X$ and $u_0= -{\it D_{n_0}}$ denotes the   material velocity   with respect to $S_0$. 
   From \eqref{keyN}, we deduce :
      \begin{equation*}
     u\,\left(1, \frac{\boldsymbol n^\star\boldsymbol F}{u}\right)=k\,\left(1,  \boldsymbol n_0^{\prime\star}\right) \quad \text{with}\quad\boldsymbol n_0^{\prime\star}=\frac{\boldsymbol n_0^{\star}}{u_0}
        \end{equation*}
    where $k$ is a coefficient of proportionality  ($k=u$).  Consequently,
     \begin{equation}
     \frac{\boldsymbol n^\star \boldsymbol F}{u}= \frac{\boldsymbol n_0^{\star}}{u_0} =\boldsymbol n_0^{\prime\star}=-  \boldsymbol w^\star  \label{key k1}
     \end{equation}
       Expression \eqref{key k1} yields $u$ and $\boldsymbol n$ as function of  $\boldsymbol w^\star$ and $\boldsymbol F$. From the knowledge of $\boldsymbol v$ we deduce $\boldsymbol n^\star \boldsymbol  v$ and consequently $ {D_n}= \boldsymbol n^\star \boldsymbol  v -u$.
         \subsection{Surfaces of discontinuity}
         We assume that  motion $\mathbf\Phi$ is a continuous function but with discontinuous derivatives on the moving surface $\Sigma$ represented in $\mathcal W_0$ by its image $\Sigma_0$ of  equation $\lambda =0$. This hypothesis implies :
         \begin{equation*}
         \boldsymbol{N}_0 ^\star\, d\boldsymbol{Z}= 0 \quad\Longrightarrow\quad \left[\frac{\partial\boldsymbol{z}}{\partial\boldsymbol{Z}}\right]\,d\boldsymbol{Z}=\boldsymbol 0
         \end{equation*}
         where brackets $\,[\ \ \, ]$  indicate the  discontinuity  jump across  $\Sigma$.     Consequently  on the surface  of discontinuity $\Sigma_0$ of equation $\lambda =0$ in $\mathcal W_0$,
      \begin{equation*}
 \{d\lambda\equiv \boldsymbol I^\star\, d\boldsymbol Z =0  \ \Longrightarrow\ [\mathcal B]\, d\boldsymbol{Z}=\boldsymbol 0\,\}\   \Longrightarrow\   \{\forall\ d\boldsymbol{X},\, [\boldsymbol w^\star]\, d\boldsymbol{X}=0\  \text{and}\  [\boldsymbol B]\, d\boldsymbol{X}= \boldsymbol 0\}
      \end{equation*}
      Then, $[\boldsymbol w^\star]=\boldsymbol 0^\star$  and $[\boldsymbol B]=\boldsymbol O$. Consequently :
      \begin{equation*}
       [\mathcal B]=\left(
       \begin{array}{cc}
      [\mu]&\ \boldsymbol 0^\star      
       \\
    \left[\mu\,\boldsymbol v\right]&\   \boldsymbol  O
       \end{array}%
       \right)
      \end{equation*}
 From \eqref{B2} and \eqref{keyF},
      \begin{equation}
      [\boldsymbol w^\star]=\boldsymbol 0^\star\quad\rm{and}\quad[\boldsymbol{v}]\,\boldsymbol{w}^\star+[\boldsymbol{F}]=\boldsymbol O\label{discontinu1}
      \end{equation}
    From \eqref{key k1},
      \begin{equation}
      \frac{\boldsymbol n^\star\boldsymbol{F}_1}{u_1}=  \frac{\boldsymbol n^\star\boldsymbol{F}_2}{u_2}=\frac{\boldsymbol n_0^\star}{u_0},\qquad \left[\frac{\boldsymbol n^\star \boldsymbol F}{u}\right]=\boldsymbol 0^\star \qquad \rm{and}\qquad[\boldsymbol{F}]=[\boldsymbol{v}]\,\frac{\boldsymbol n_0^\star}{u_0}\label{discontinu2}
      \end{equation}
    where subscripts $1$ and $2$ indicate the upstream and downstream values  of  the discontinuity surface $\Sigma$. 
    It is possible to calculate the discontinuities of each tensor, image in $\mathcal W$ of a tensor defined on $\mathcal W_0$.\\ 
      We can interpret  \eqref{discontinu2}: 
      applying $d\boldsymbol{X}$ to \eqref{discontinu2}$^1$, we obtain :
      \begin{equation*}
      \frac{\boldsymbol n^\star d\boldsymbol{x}_1}{u_1}=  \frac{\boldsymbol n^\star d\boldsymbol{x}_2}{u_2}
      \end{equation*}
      The projections on $\boldsymbol n$ of vectors $d\boldsymbol{x}_1= \boldsymbol{F}_1\,d\boldsymbol{X}$ and 
        $d\boldsymbol{x}_2= \boldsymbol{F}_2\,d\boldsymbol{X}$ are proportional to $u_1$ and $u_2$, respectively.\\
      Equation\eqref{discontinu1}$^2$ yields $[\boldsymbol{F}]=-[\boldsymbol{v}]\,\boldsymbol{w}^\star$; then $d_2\boldsymbol{x}-d_1\boldsymbol{x}= -[\boldsymbol{v}]\,\boldsymbol w^\star d\boldsymbol{X}$ and the direction of vector $d_2\boldsymbol{x}-d_1\boldsymbol{x}$  corresponds   to the direction of $[\boldsymbol{v}]$.\\ Let us notice that   $[\boldsymbol{v}]$ is not necessary normal to the surface of discontinuity.
      \section{Shock waves}
      When $u\neq 0$,
  surfaces of discontinuity are  shock waves \cite{Hadamard,Courant }.
      \subsection{Mass conservation}
       The reference mass density  being given in $\mathcal W_0^\prime$ and the mass  conservation corresponding to
      $
      \rho\ {\rm det}\,\boldsymbol F =f(\boldsymbol X) 
      $, we get
     across a shock wave in $\mathcal W_0^\prime$ : 
  \begin{equation*}
  \left[f(\boldsymbol X)\right]=0
   \end{equation*}
By using   (\footnote{From Relation \eqref{discontinu2}, we get :\\ 
$\displaystyle [\boldsymbol{F}]=[\boldsymbol{v}]\,\frac{\boldsymbol n_0^\star}{u_0} \quad\Longrightarrow\quad \boldsymbol{F}_2\boldsymbol{F}_1^{-1}=\boldsymbol 1+[\boldsymbol{v}]\,\frac{\boldsymbol n_0^\star}{u_0}\,\boldsymbol{F}_1^{-1}\quad\Longrightarrow\quad\boldsymbol{F}_2\boldsymbol{F}_1^{-1}=\boldsymbol 1+[\boldsymbol{v}]\,\frac{\boldsymbol n^\star}{u_1}$.\\
      From $\det  (\boldsymbol 1+\boldsymbol K\,\boldsymbol L^\star)= 1+\boldsymbol L^\star\, \boldsymbol K$, where $\boldsymbol L$ and $\boldsymbol K$ are two \textit{3-D} vectors, we get : 
     \begin{equation*}
      \frac{\det \,\boldsymbol{F}_2}{\det \,\boldsymbol{F}_1}= 1+\frac{[u]}{u_1}=\frac{u_2}{u_1}\quad\Longrightarrow\quad \frac{\det \,\boldsymbol{F}_i}{u_i}= \frac{[\det \,\boldsymbol{F}]}{[u]},\quad i\in\{1,2\}. 
      \end{equation*}
  }), one deduces :
   \begin{equation*}
      [\rho]=f(\boldsymbol X) \left[\frac{1}{\text{det}\, \boldsymbol F}\right]=-f(\boldsymbol X)\frac{[{\text{det}\, \boldsymbol F}]}{\text{det}\, \boldsymbol F_1\  \text{det}\, \boldsymbol F_2}=-f(\boldsymbol X)\frac{[u]}{u_2\; \text{det}\, \boldsymbol  F_1}  
       \end{equation*}
      Then, $ \displaystyle [\rho]=- \frac{[u]}{u_2}\,\rho_1, $     which  implies the  geometrical property (which is  not related with Hamilton's principle) : 
      \begin{equation}
      [\rho\, u]=0\label{masscons} 
      \end{equation}  
      \subsection{Hamilton's principle}
      When $  \mathcal F^\star -{\rm{Div}}\ T=\boldsymbol{ 0}^\star$ and $ \mathcal F_0^\star -{\rm{Div}}_0\ T_0=\boldsymbol{ 0}^\star$ corresponding to  conservative motion equations (see Section 3.2), the variation of Hamilton's action writes :
      \begin{equation*}
   \delta a=\int_{\Sigma}\boldsymbol N^\star \left[T\right]\,\boldsymbol{\tilde\zeta}\, d\sigma = \int_{\Sigma_0}\boldsymbol N_0^\star \left[T_0\right]\,\boldsymbol{\hat\zeta}\, d\sigma_0
      \end{equation*}
       \subsubsection{First variation}
      From Hamilton's principle, we obtain :
      \begin{equation*}
  N^\star \left[T\right] =\boldsymbol{ 0}^\star
      \end{equation*}
        \begin{equation}
    \left[D_n\, p+(e+p) u\right]  =0\qquad\text{and}\qquad\left[\rho u\, \boldsymbol v+ p\,\boldsymbol n\right]=\boldsymbol{ 0} \label{CRH1}
      \end{equation}
   Equation \eqref{CRH1}$^2$ corresponds to the  \textit{conservation of impulsion} across the shock wave. We can    write :
   \begin{equation*}
   \left[\rho u\, \boldsymbol v+ p\,\boldsymbol n\,  \right]=\boldsymbol 0\quad\Longleftrightarrow\quad
    \left\{
   \begin{array}{l}
   \displaystyle  [p+\rho\, 
   {u}^2]  =0  \\ \\
    \left[\boldsymbol v_{tg}\right]=\boldsymbol 0
   \end{array}%
   \right.
   \end{equation*} 
   where $ \boldsymbol v_{tg}$ is the tangential velocity component    at the shock wave. Equivalently, we obtain : 
    \begin{equation*}
    \left[\boldsymbol v\right] =[u]\,\boldsymbol n\label{nvel}
     \end{equation*}
   From \eqref{discontinu1}$^2$ we additively obtain :
 \begin{equation}
   [F] =-[u]\,\boldsymbol n\, \boldsymbol w^\star=  \frac{[u]}{u_0}\, \boldsymbol n\,\boldsymbol n_0^\star\label{disFF}
 \end{equation} 
  From   \eqref{masscons} and $ [p+\rho\, 
   {u}^2]  =0$,   \eqref{CRH1}$^1$ is equivalent to $\displaystyle\left[\frac{1}{2}\, \boldsymbol v^2+h -D_n\,u\right]=0$ and $\left[\boldsymbol v_{tg}\right]=\boldsymbol 0$ implies $\left[\boldsymbol v^2\right]=\left[u^2+2D_n\, u\right]$. Consequently,
   \begin{equation*}
    \displaystyle\left[\frac{1}{2}\,u^2+h\right]=0
   \end{equation*}
   which corresponds to the conservation of energy.
   We can resume the conditions associated with the first variation corresponding to $\tilde {\boldsymbol \zeta}$ :
   \begin{equation}
   [\rho\,u]=0, \qquad  [p+\rho\, 
   {u}^2]  =0,\qquad \left[\boldsymbol v\right] =[u]\,\boldsymbol n,\qquad\left[\frac{1}{2}\,u^2+h\right]=0\label{RHC1}
   \end{equation}
  Equations \eqref{RHC1} represent the Rankine-Hugoniot conditions for fluid shock--waves in the  space--time $\mathcal W$, where we can add geometrical condition \eqref{disFF}.
      \subsubsection{Second variation}
       We assume the shock wave $\Sigma_0$ is represented by equation $\lambda =0$ in $\mathcal W_0$. 
     The mapping $\mathbf\Phi$ being continuous along the shock wave, we obtain again \eqref{discontinu1}.\\
      The principle of Hamilton implies :    \begin{equation*}
      N_0^\star \left[T_0\right] \equiv\boldsymbol{I}^\star [{T_0}]=\boldsymbol{ 0}^\star 
         \end{equation*} 
      From the value of ${T_0}$ obtained in Section 3.2, we obtain :
      \begin{equation}
     \left[\boldsymbol v^\star F  + m\, \boldsymbol w^\star\right] =\boldsymbol 0^\star \label{SCR}
      \end{equation}
      which is the only shock condition  associated with the virtual displacement $\boldsymbol {\hat\zeta}$. Equation \eqref{SCR} corresponds to three scalar equations  through the shock wave $\Sigma_0$.
      From \eqref{discontinu1}$^2$, \eqref{discontinu2}$^2$ and
      \eqref{SCR}, we respectively deduce :
     \begin{equation*}
     \boldsymbol w^\star\, d\boldsymbol X = 0 \quad\Longrightarrow\quad [F]\,d\boldsymbol X = \boldsymbol 0\quad  \text{and}\quad [\boldsymbol v^\star F]\,d\boldsymbol X=0
     \end{equation*}
      or
       \begin{equation*}
       \boldsymbol w^\star\, d\boldsymbol X = 0 \quad\Longrightarrow\quad d_1\boldsymbol x=d_2\boldsymbol x\quad  \text{and}\quad  \boldsymbol v^\star_1\,d_1\boldsymbol x=\boldsymbol v^\star_2\,d_2\boldsymbol x 
       \end{equation*}
       From \eqref{key k1}, 
      \begin{equation*}
      \boldsymbol w^\star\, d\boldsymbol X = 0\quad\Longleftrightarrow\quad\boldsymbol n^\star\, d\boldsymbol x = 0
      \end{equation*}
      Consequently,
      \begin{equation*}
  \boldsymbol n^\star\, d\boldsymbol x = 0\quad\Longrightarrow\quad \left[\boldsymbol v^\star\right]\,\, d\boldsymbol x=0
      \end{equation*}
and there exists a Lagrange multiplier $\alpha$ such that   $\left[\boldsymbol v\right] =  \alpha\,\boldsymbol n$; consequently the discontinuity of $\boldsymbol v$ is normal to the shock wave.\\ Due to $\alpha = \left[\boldsymbol n^\star\, \boldsymbol v\right] = [u]$, we deduce :
 \begin{equation}
\left[  \boldsymbol v\right] =[u]\, \boldsymbol n \quad \rm{and}\quad \left[\boldsymbol v_{tg}\right]=\boldsymbol 0\label{vdis}
 \end{equation}
 and we obtain again \eqref{disFF}.\\
 From $[\Omega]=0$, we get $[m\,\boldsymbol w^\star]=\displaystyle\left[\frac{1}{2}\,\boldsymbol v^2-h\right]\, \boldsymbol w^\star$. 
 Moreover, $\left[\boldsymbol v^\star\,F\right]=
  \boldsymbol v_1^\star\,[F]+\left[\boldsymbol v^\star\right]  F_2= \boldsymbol v_1^\star\,[F]+[u]\,\boldsymbol n^\star F_2\,$.\\
   From $\boldsymbol n^\star\,\boldsymbol v_1= u_1+D_n$ and
  because  \eqref{discontinu2}$^2$ and \eqref{vdis}, $[F]=[\boldsymbol v]\, \boldsymbol n_0^{\prime\star}=[u]\, \boldsymbol n\,\boldsymbol n_0^{\prime\star}$ and from \eqref{discontinu2}$^1$, we obtain $ \boldsymbol n^\star F_2 = u_2\, \boldsymbol n_0^{\prime\star}$.
   We deduce :
   \begin{equation*}
   \left[\boldsymbol v^\star F\right] =\left[u\right] \boldsymbol v_1^\star\,\boldsymbol n\,\boldsymbol n_0^{\prime\star}+\left[u\right]\, u_2\, \boldsymbol n_0^{\prime\star}
 =[u] \left(u_1+D_n+u_2\right)   \boldsymbol n_0^{\prime\star}=-\left[u^2+D_n\,u\right]
 \boldsymbol w^\star
 \end{equation*} 
 Consequently,
  $[\boldsymbol v^\star F+m\,\boldsymbol w^\star ]=- \left[u^2+D_n\,u+h-\displaystyle\frac{1}{2}\,\boldsymbol v^2\right]\boldsymbol w^\star =\boldsymbol 0^\star$ and finally, 
  \begin{equation*}
\left[\frac{1}{2}\,u^2+h\right]  \,\boldsymbol w^\star=\boldsymbol 0^\star\quad\Longleftrightarrow\quad\left[\frac{1}{2}\,u^2+h\right]=0 \label{disener}
\end{equation*} 
 We can resume the conditions associated with the second variation :
 \begin{equation}
 [\rho\, 
 {u}]  =0,\quad  \left[\boldsymbol v\right] =[u]\,\boldsymbol n,\quad \left[\frac{1}{2}\,u^2+h\right]=0\label{CRH2}
 \end{equation}
 Equations \eqref{CRH2} represent the Rankine--Hugoniot conditions for fluid shock wave in $\mathcal W_0$. 
\section{Conclusion}Only the first variation of Hamilton's action yields  the well--known Rankine--Hugoniot conditions   for fluids. The second variation yields only a part of the Rankine--Hugoniot conditions by missing the condition $[p+\rho\, 
{u}^2]  =0$. \\  
This difference of results can be explained : in the \textit{4-D} space-time, along surface $\Sigma$, the test function (or virtual displacement)  $\boldsymbol {\tilde \zeta}$ is a continuous function.   But $\mathcal B$ is discontinuous through the shock wave and due to \eqref{relvirtualdsipl} the virtual displacement $\boldsymbol {\hat \zeta}$ is discontinuous;  consequently we cannot apply the fundamental lemma of variation calculus which must be used with continuous test functions. Then, only the first variation associated with $\boldsymbol {\tilde \zeta}$  in the \textit{4-D} space-time is able to get the whole Rankine--Hugoniot conditions on  shock waves.\\  It is noticeable that the variation ${\tilde\delta}$ gives the conservation of  energy, while the   variation ${\hat\delta}$ does not give the conservation of entropy. This fact was the subject of   strong discussions in the shock waves' studies and Hamilton's principle directly yields  the result without any ambiguity. The problem of  Lax's condition for entropy is a different problem relevant of the second law of thermodynamics \cite{Lax,Liu,Liu-Ruggeri}.  
\\
 To conclude, the Hamilton action in   the \textit{4-D} space--time  is a powerful tool to obtain  shock conditions in multi-dimensional spaces. It unambiguously allows us to obtain the equations already known and to consider the study of more complex cases not yet considered in the literature \cite{Gavrilyuk2}.  The forthcoming article \cite{Gavrilyuk2} also uses Hamilton's principle and obtain complementary relations to Rankine--Hugoniot conditions   for second--gradient or bubbly fluids. In this paper devoted to classical fluids, we can nevertheless observe the importance of the virtual displacements in the \textit{4-D} space--time to obtain the correct Rankine--Hugoniot conditions; this is an important difference with  the virtual displacements in the \textit{4-D} reference--space associated with Lagrange variables. Another important observation is that the  virtual displacements in the \textit{4-D} space--time  naturally yield the conservation of energy through a shock wave. 
However, the Hamilton action cannot be directly used for hyperbolic dissipative systems; in such cases, the principle of virtual power in space--time may be a useful extension of the Hamilton principle as it was considered in \cite{Germain}.\\

 \noindent {\footnotesize{\textbf{Acknowledgments :}
{The paper is   dedicated  to Professors Masaru Sugiyama and Giuseppe Toscani on the occasion of their 70 years.
		This work was partially supported by National Group of Mathematical Physics
		GNFM-INdAM (Italy).}

\end{document}